\documentclass[%
reprint,
showpacs,
preprintnumbers,
amsmath,amssymb,
aps,
]{revtex4-1}

\usepackage{graphicx}
\graphicspath{ {./figures/} }
\usepackage{dcolumn}
\usepackage{bm}
\usepackage[T1]{fontenc}
\usepackage{textcomp}

\usepackage{amsfonts}
\usepackage{upgreek}     
\usepackage[T1]{fontenc} 
\usepackage{ae,aecompl}  

\newcommand{\micron}{~\ensuremath{\upmu\text{m}}}
\newcommand{\umns}{\ensuremath{\upmu\text{m}}}
\newcommand{\microsec}{~\ensuremath{\upmu\text{s}}}
\newcommand{\atpo}{~\ensuremath{a^3\Pi_1}}

\begin{document}
\title{Imaging Cold Molecules on a Chip}
\author{S. Marx}
\author{D. Adu Smith}
\author{M. J. Abel}
\author{T. Zehentbauer}
\author{G. Meijer}
\author{G. Santambrogio}
\email{gabriele.santambrogio@fhi-berlin.mpg.de}
\affiliation{Fritz-Haber-Institut der Max-Planck-Gesellschaft,
Faradayweg 4-6, 14195 Berlin, Germany}
\date{\today}

\begin{abstract}
\noindent We present the integrated imaging of cold molecules in a
microchip environment. The on-chip detection is based on REMPI, which is
quantum-state-selective and generally applicable. We demonstrate and characterize
time-resolved spatial imaging and subsequently use it to
analyze the effect of a phase-space manipulation sequence aimed at
compressing the velocity distribution of a molecular ensemble
with a view to future high-resolution spectroscopic studies. The
realization of such on-chip measurements adds the final fundamental
component to the molecule chip, offering a new and promising route for
investigating cold molecules.  
\end{abstract}

\maketitle


The idea of integrating various parts of a laboratory on a microchip
is now over twenty years old and was motivated by a desire for
simplification, reduction of costs, 
portability, speed of measurement, and ease of reproducibility. A
reduced size allows for shorter transport times as well as for 
large field gradients, and therefore strong confinement forces
created at low field strengths, employing only modest
voltages or current densities. For
physics, the atom chip~\cite{Reichel:AtomChip} 
and ion chip~\cite{Stick2006:p36} have been employed in fields as
diverse as quantum computation~\cite{Ospelkaus2011:p181}, many-body
non-equilibrium physics~\cite{Gring2012:p1318}, and gravitational
sensing~\cite{Zoest2010p1540}. For chemistry, the lab-on-a-chip
shrinks the pipettes, beakers and test tubes of a modern lab onto a
microchip-sized substrate~\cite{Daw:2006p2231}, with applications from
the international space station~\cite{Morris:2012p830} to
anti-terrorism~\cite{Frisk2006:p1504}.  The molecule
chip~\cite{Meek:2009p1699,Zeppenfeld_Nature491p570_2012}, however, is currently in its infancy, but
promises a marriage between fundamental quantum physics
and the richness of the chemical world. A particular advantage of
using molecules instead of atoms on a chip is that they can be coupled
to photons over a wider range of frequencies by their rotational
and vibrational degrees of freedom. Moreover, for chemists the
molecule chip offers the prospect of extending the control of
molecular concentrations and interactions to the level of single
molecules with the accuracy in interaction energy enhanced to the mK
level or beyond.

A major obstacle that has delayed the development of the molecule chip
arises from this same richness of molecules' internal degrees of
freedom. The complicated level structures of molecules result in a general
lack of closed two-level systems that are necessary for efficient laser 
cooling and detection using absorption or laser-induced fluorescence. 
In addition, in the presence of a physical structure such as a microchip, 
scattering or laser-induced fluorescence from surfaces adds noise to 
images, something which is critical when working with small samples. 
For these reasons, molecule detection has always been carried out 
tens of centimeters away from the chip.  In recent years, we have
demonstrated that one can exploit the cooling provided by a supersonic
expansion by loading the chip with cold molecules directly from a
molecular beam.\cite{Meek:2008p153003}
The molecule chip has been used to trap, decelerate and accelerate polar
molecules~\cite{Meek:2009p1699}, as well as manipulate both their 
rotational~\cite{Santambrogio:2011p1674} and
vibrational~\cite{Abel:2012p2225} states, using microwave and infrared
radiation, respectively. This latter function can be
used both as a spectroscopic tool and as a way to prepare a sample in
a desired quantum state. Here we show on-chip molecule detection,
adding the final fundamental component to the molecule chip. 
Our detection is based on resonant-enhanced multi-photon ionization
(REMPI)~\cite{Ashfold1994:p57}, which we have already successfully
employed in the past for off-chip
detection.\cite{Santambrogio:2011p1674,Abel:2012p2225,Meek:2011p033413}
REMPI is quantum state selective, can
be saturated with a few mJ/mm$^2$ of laser light for most molecules,
is intrinsically background-free, and is of general applicability.
While in the simplest implementation of REMPI
one would simply count the ions, we take the further step of
using ion optics to create a time-resolved spatial image of the molecules.

\begin{figure}
\centering
\includegraphics[width=0.45\textwidth]{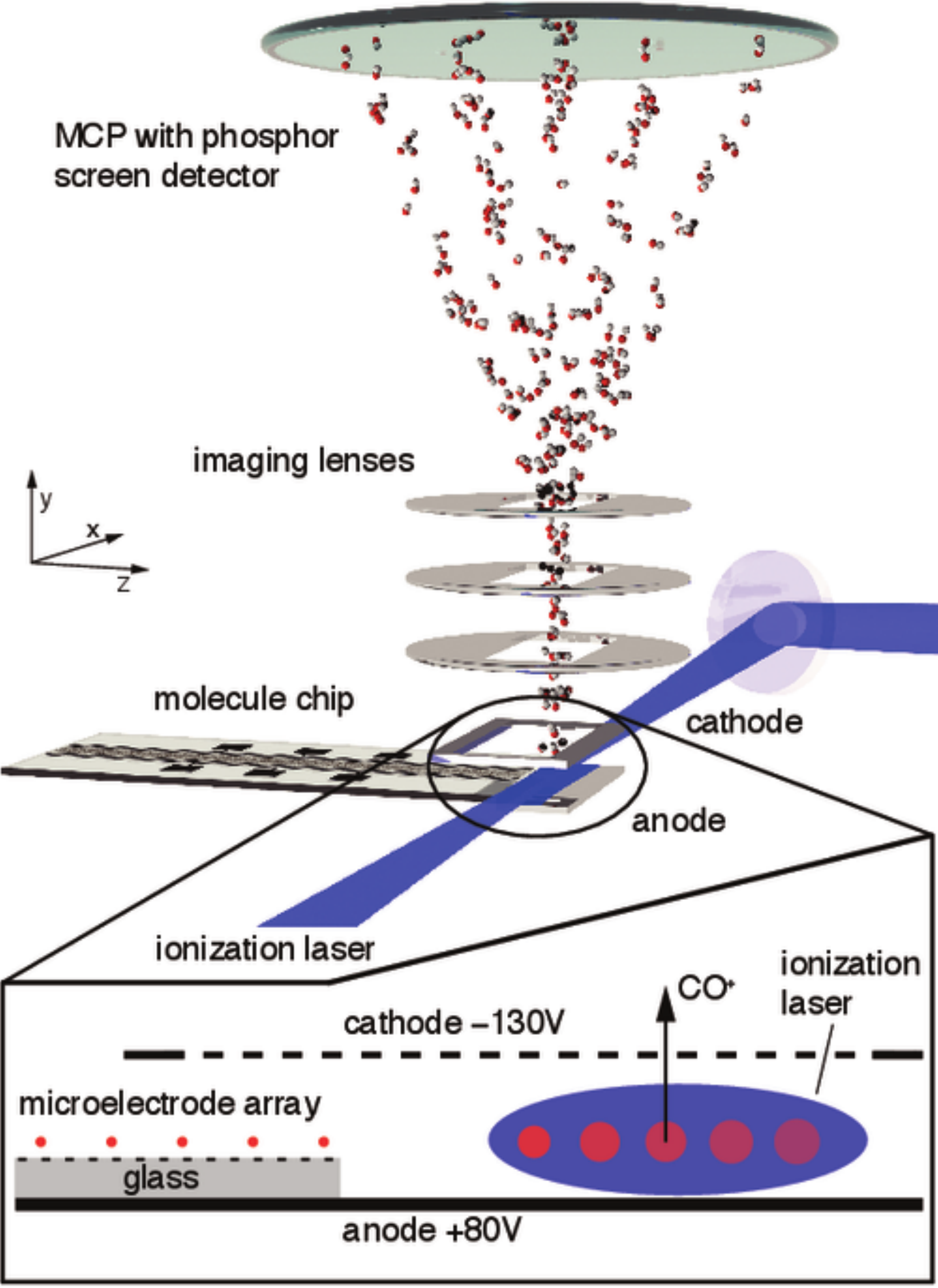}
\caption{Metastable CO molecules
 on the molecule chip are ionized via REMPI using a pulsed 283~nm laser. The
 detection region is composed of two electrodes: an anode and a
 cathode. The cathode is a rectangular ring. Together the two
 electrodes protect the cations
 from stray fields in the region of the chip. Ion optics are then used to form
 an image on a MCP detector backed by a phosphor screen. A CCD camera
 (not shown) is used to record the image of the molecules from the
 phosphor screen. A sketch of the detection region is shown in the
 inset, not to scale. Two points in time are shown: on the left, when the 
 molecules are still trapped above the microelectrode array; on the right, after
 ballistic flight, upon reaching the axis of the imaging lenses. 
\label{fig:Setup}} 
\end{figure}

We load the molecule chip directly from a molecular beam, which is
generated by expanding a mixture of 20\% $^{13}$CO in Kr, cooled to
140~K, through a pulsed solenoid valve operated at 10~Hz. The
expansion is skimmed to form a collimated beam with a mean velocity
of 330~m/s and intersected at right angles by a 206~nm laser pulse~\cite{Velarde_RevSciInstr81p063106_2010}
exciting the molecules from the ground electronic state to the upper
$\Lambda$-doublet component of the $a^3\Pi_1$, $v=0$, $J=1$ state. 
This metastable state has a lifetime of
2.6~ms~\cite{Gilijamse:2007p1477} and consists of two 
low-field seeking sub-levels and a third sub-level with no significant
Stark shift. 

Our molecule chip is an array of gold microelectrodes on glass
substrate.~\cite{Meek:2009p055024} 
By sinusoidally modulating the voltages applied to these
electrodes, tubular microtraps that travel smoothly along the chip in
the direction parallel to the molecular beam $z$ can be produced. 
The speed of the traveling traps is proportional to the frequency of the
modulation. For example, a waveform of 2.75~MHz frequency will move
the traps at 330~m/s. Once the molecules are trapped they can be
decelerated to a standstill by chirping down the frequency of the applied
waveforms.~\cite{Meek:2009p1699} In all the experiments presented here, molecules are
immediately decelerated to 138~m/s upon arrival on the chip: this
deceleration is sufficient to separate the trapped molecules from the
background of untrapped molecules. 
The depth of the microtraps depends on the amplitude of the waveforms
applied to the microelectrodes and on the acceleration of the traps
(chirp rate of the waveforms), but not on the velocity of the traps
(frequency of the waveforms).

For detection, the molecules are ionized using (1+1) REMPI via the $b^3\Sigma^+$,
$v=0$, $N=1$ state using 0.8~mJ/mm$^2$ of laser light at 
283~nm~\footnote{Light at 283~nm is obtained by frequency doubling the
output of a NarrowScan from Radiant Dyes Laser, pumped by the second harmonic of a
Nd:YAG laser.} that propagates parallel to the chip surface.
For the simplest implementation of REMPI (i.e.~no imaging) we ionize
the molecules 
in the region of the microtraps after switching the traps off. All
microelectrodes are switched to 0~V and, additionally, a rectangular
ring electrode parallel to the chip surface, but offset above the chip
by 4~mm, is held at $-$100~V~\footnote{The
  exact voltage applied to 
  the rectangular ring electrode for non-imaging detection is not
  critical.}. This way the 
microelectrode array acts as an anode for the newly generated cations
and the rectangular ring as a cathode. After flying through the
cathode, the cations are then collected by a microchannel plate (MCP)
detector.  

The electric field homogeneity achievable above the microelectrode
array, however, is not sufficient for spatial imaging. Therefore, we
added an extra anode directly behind the microtraps region
(Fig.~\ref{fig:Setup}), wider than 
the microtraps in the $x$ direction and 10~mm long in the $z$
direction. We hold this anode at $+$80~V. The molecules expand
ballistically while they fly above the anode after being ejected from
the microtraps. Hence this electrode is recessed by 2~mm under the
plane of the traps to leave enough space for the expansion. As a cathode
we use the same rectangular ring as for the non-imaging detection. For
imaging, the cathode is held at $-$130~V. Together the anode and
cathode reduce the inhomogeneity of the electric field and therefore
any associated imaging aberrations. The ions generated by REMPI are imaged
using ion optics onto a MCP detector with phosphor screen situated
40~cm above the chip surface and recorded using a fast CCD camera.  

\begin{figure}
\centering
\includegraphics[width=0.45\textwidth]{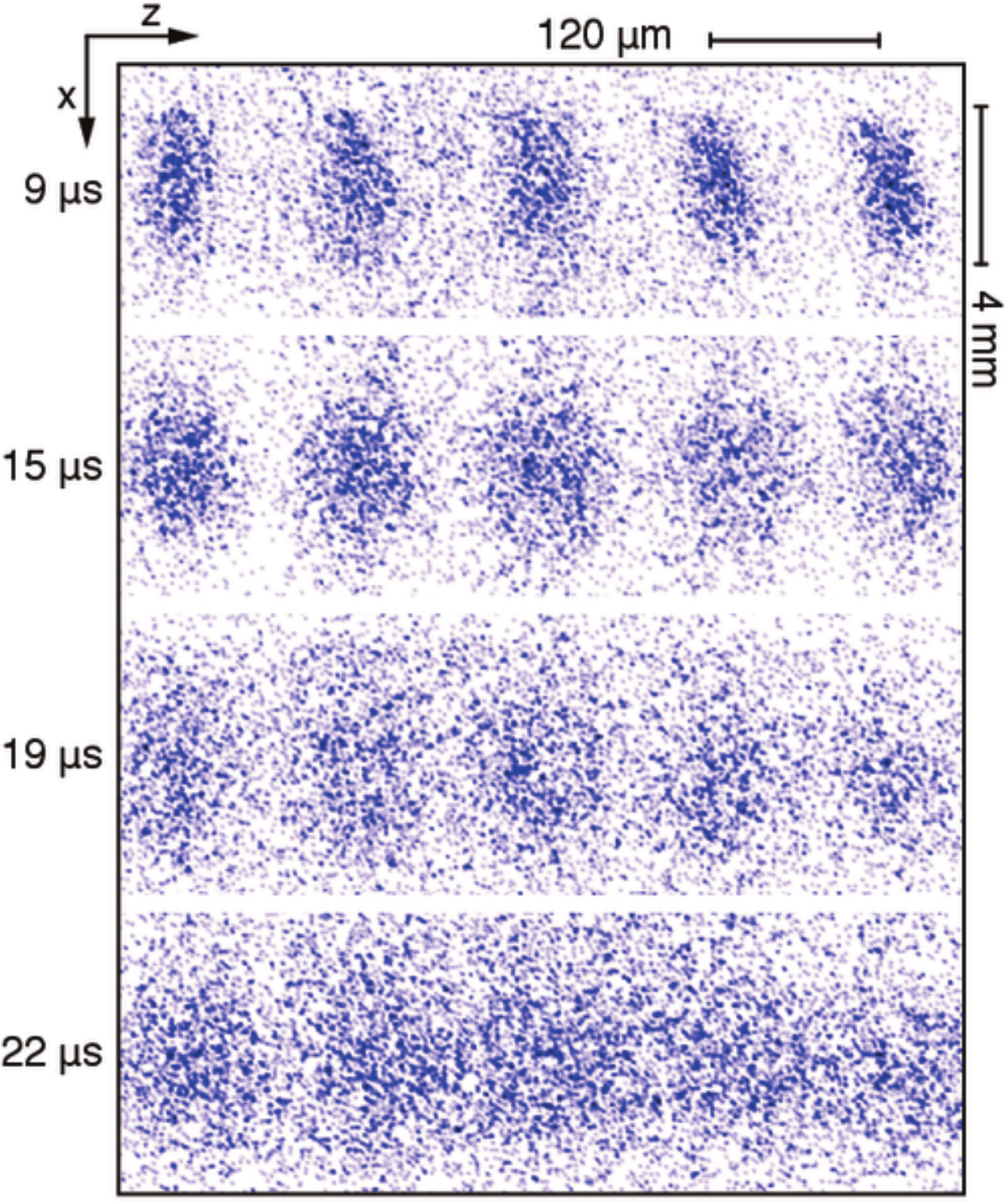}
\caption{Two-dimensional spatial imaging of CO
   molecules.  Molecules are imaged for different ballistic expansion
 times. The vertical direction is along the length of the
 microtraps, $x$. The extreme aspect ratio of the microtraps, $x:z=200:1$, is
 reduced to approximately $7:1$ by the asymmetric magnification factors of the
 lenses. Some ion-optics aberrations can be observed on the right hand
 side of the image, in particular at shorter times. The experimental
 procedure runs at a rate of 10 Hz and the images are the sum of
 approximately 10$^5$ experimental cycles. We record approximately
 1--2~counts/s. The resolution of the imaging system in the $z$
 direction is 0.78 pixel/\umns.\label{fig:2Dclouds}}
\end{figure}

\begin{figure}
\centering
\includegraphics[width=.45\textwidth]{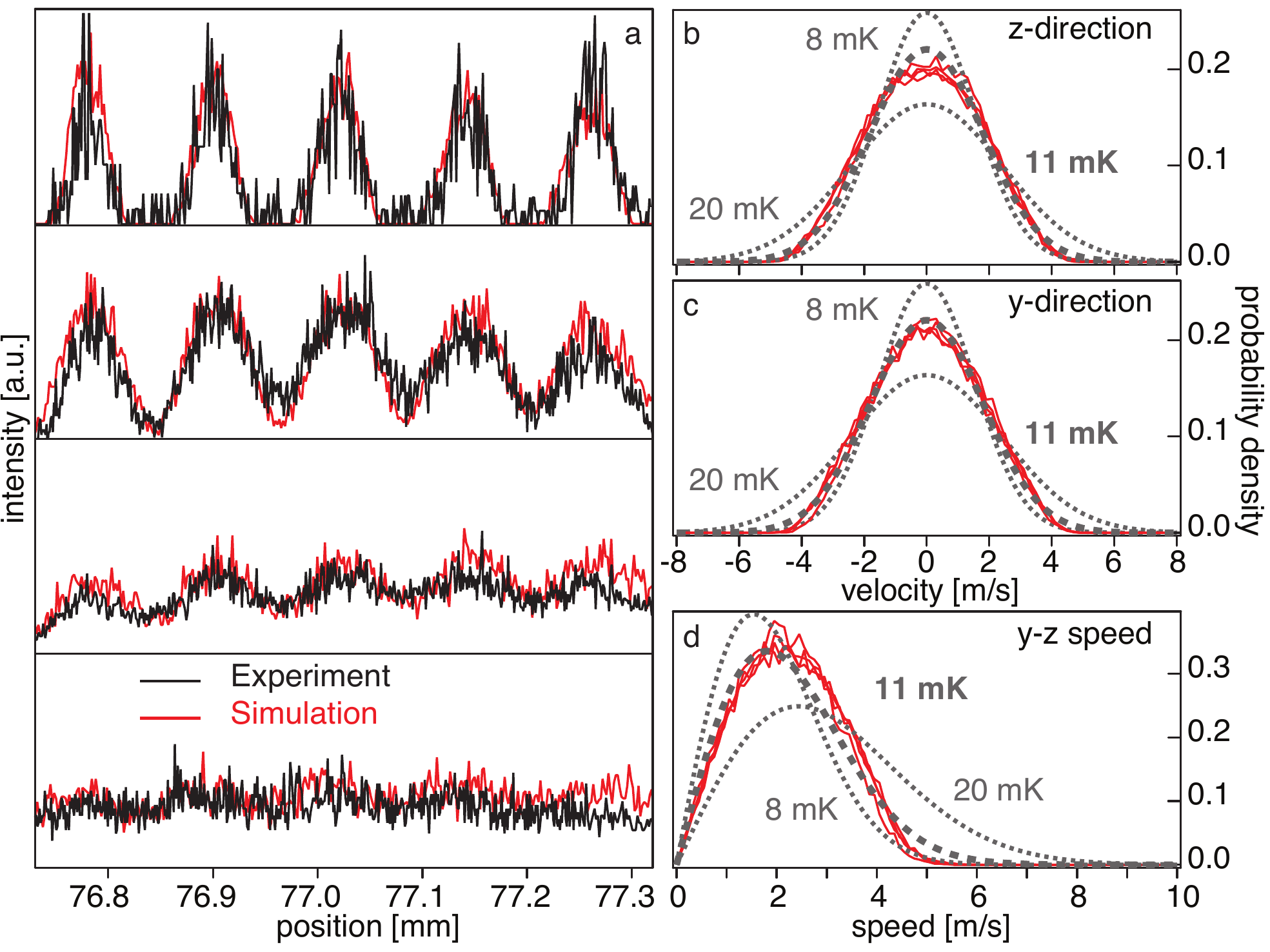}
\caption{(a) Integrated signals measured under the same conditions as
  in Figure~\ref{fig:2Dclouds}, together with trajectory simulations. 
 The experimental signals are each scaled to the (unscaled) simulations, which, for each
 expansion time, were run with the same number of input molecules.
 (b) and (c),  Velocity distributions in the $z$ and $y$ directions from
 trajectory simulations (four almost overlapping solid red lines). Results are shown
 for all four sets of data in (a). For comparison, calculated 
 Maxwell-Boltzmann distributions are plotted for 8~mK and 20~mK
 (dotted lines) and 11~mK (dashed line), which is the best numerical fit. 
 (d) Speed distributions in the $y$-$z$ plane from trajectory simulations
 (solid red lines) and corresponding Maxwell-Boltzmann distributions.
 While the simulated velocity distributions appear well approximated by thermal 
 distributions, the deviation of the speed distribution indicates that there is some
 correlation between the velocities in the $y$ and $z$ directions.
 \label{fig:Time}} 
\end{figure}

The ion optics for our experiment are a standard set of three
asymmetric ion lenses located 26~mm from the
chip surface, spaced by 10~mm and all
held at $-400$~V. The openings in the centers
of the ion lenses of 22 by 72~mm provide magnification factors 
at the MCP of 2 and 58, in the direction of the tubular trap axis and
the direction of the trap movement, respectively. We chose this
asymmetric magnification because of the large aspect ratio of the
tubular microtraps (200). Four images are shown in
Figure~\ref{fig:2Dclouds}. They show molecules ejected from the
microtraps toward the detection region, after a ballistic expansion of
9, 15, 19, and 22\microsec. One can clearly distinguish the 
single traps and follow the ballistic expansion of the molecular
clouds. At a still relatively short time of flight of
22\microsec\ the resolution of the individual microtraps is almost
completely washed out. This highlights the importance of on-chip
detection for spatial imaging. The REMPI laser fires when the
molecules are on the axis of the imaging lenses. With the distance
between release and detection being fixed, the time for which
molecules are in ballistic flight is given by the center of mass
velocity of the molecular ensemble. We control this velocity by
choosing the speed of the microtraps at the point at which the molecules are
released, i.e. a lower microtraps velocity gives a longer time for
ballistic flight. For the ballistic expansion images shown in Fig.~\ref{fig:2Dclouds},
the molecules were released when traveling uniformly at 336, 207, 162,
and 138~m/s, respectively, for the expansion times of 9, 15, 19, and
22\microsec\ to cover the 3-mm distance between the point of release from the
microtraps and the axis of the imaging lenses in the detection region. Each microtrap
releases its molecules upon reaching the end of the microelectrode
array, i.e. the clouds are released sequentially, each time at the
same position.~\footnote{Numerical simulations show that each trap
  rapidly opens out upon arrival at the end of the microelectrode
  array. This happens within hundreds of nanoseconds,
  i.e. instantaneously for the molecules. The fields disappear with a
  monotonic decrease in the electric field strength gradient.}  
The expansion time is given for the central cloud of each image, whereas 
the clouds at either side expand for slightly shorter or longer times. 
With the signal-to-noise ratio of the data presented here, the small
differences in the cloud widths within the same image, due to this difference 
in expansion time, are masked by the experimental noise (see also Fig.~\ref{fig:Time}).
The purpose of recording measurements at different times is to follow
the evolution of the system. This, of course, is only meaningful if
the initial conditions are the same for each measurement. We therefore
made sure that the molecules experienced the same trap depth and shape
for every measurement. We achieved different microtraps velocities at
the point of release by applying a fixed acceleration of 10$^6$~m/s$^2$
for different time durations. A waveform amplitude of 160~V
(peak-to-peak) applied to the microelectrodes creates 28~mK deep traps
for CO molecules in low-field-seeking components of the \atpo, $v=0$, $J=1$ level,
under an acceleration of 10$^6$~m/s$^2$. Upon reaching the
desired final velocity, the motion is made uniform and this increases
the trap depth to 57~mK while the trap diameter increases from 4 to 20\micron.

The dynamics of the molecules along the 4-mm length of the microtraps ($x$ direction)
is negligible for the experiments presented here because the molecules 
almost never experience a force in that direction during the 
relatively short time they spend on the chip. 
We therefore integrate the signal along the $x$ direction (vertical axis
of the images) and concentrate on the perpendicular direction.
The results are shown in Fig.~\ref{fig:Time}a together with the results
of numerical trajectory simulations. Experiment and simulations show good agreement,
demonstrating that both the ballistic expansion and the ion imaging
process are not significantly hindered by stray electric fields. 

\begin{figure}
\centering
\includegraphics[width=.45\textwidth]{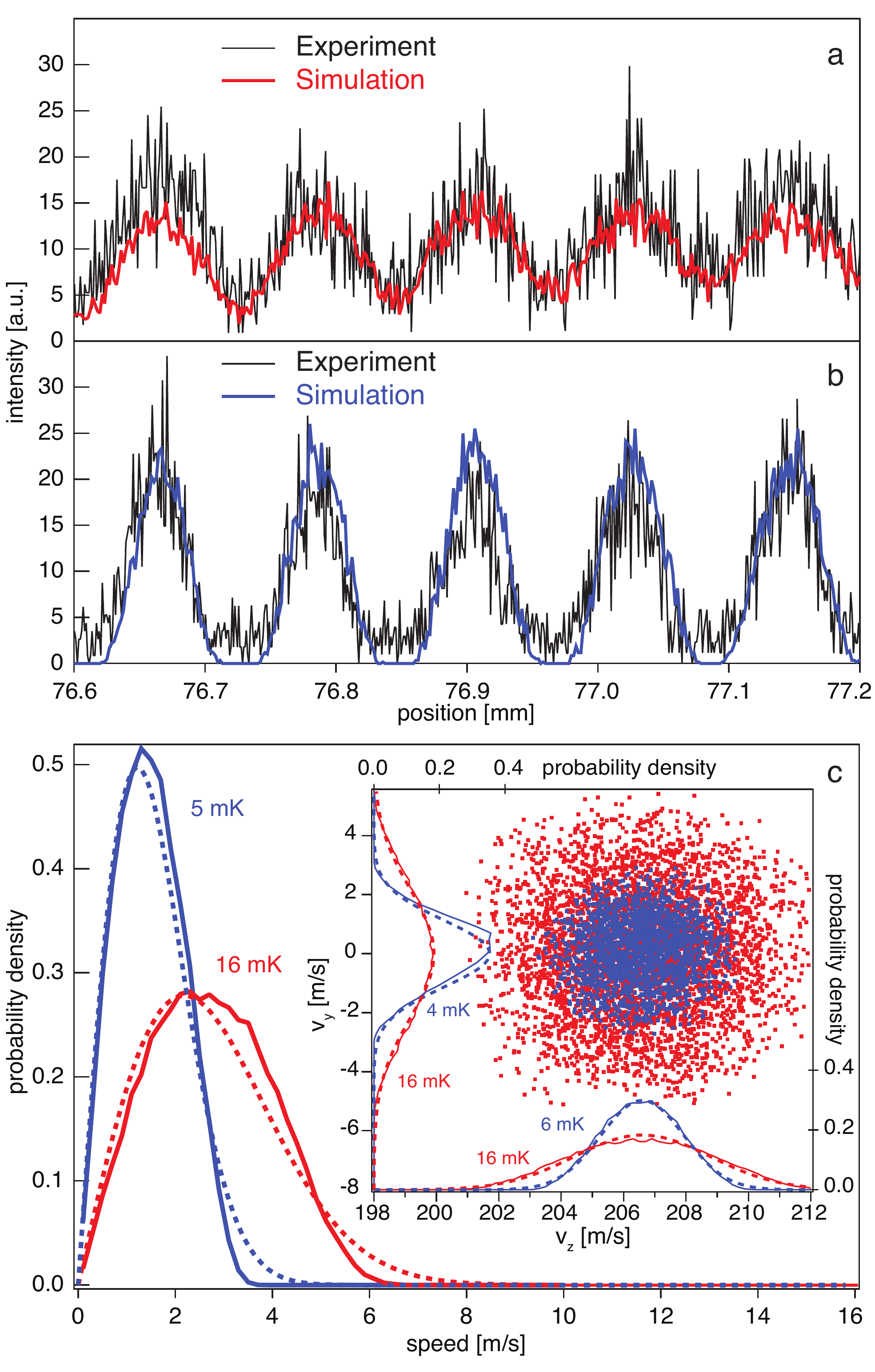}
\caption{ Integrated signals
 for molecules guided in traps with constant depth (a) and decreasing depth (b),
 together with results from trajectory simulations. There is no
 rescaling between (a) and (b). (c), Speed
 distribution in the $y$-$z$ plane for the same conditions as in
 (a) and (b). The inset in (c) shows the two-dimensional velocity
 distribution, with projections in both directions. The distributions
 in the $y$ and $z$ directions after phase-space manipulation are different,
 showing that the phase-space manipulation does not act
 isotropically. Results from trajectory simulations are shown with
 solid lines, while Maxwell-Boltzmann distributions for the best fit
 temperatures are shown as dashed lines. Simulations show a number loss of
 approximately 5\% during the phase-space manipulation. 
\label{fig:cooling}} 
\end{figure}

We have a molecular density of 10$^7$/cm$^3$ in the traps, which
corresponds to about five molecules per microtrap. The sensitivity of
our REMPI detection allows us to work
under such conditions. Indeed, analysis of the trajectory simulations
shows that this trapped ensemble 
of molecules does not have a perfect thermal distribution.
However, a comparison of the molecular velocities to a Maxwell-Boltzmann
distribution remains helpful for understanding the order of magnitudes of
the observed phenomena. In Figs.~\ref{fig:Time}(b)-(d), the
calculated velocity distributions in the $z$ and $y$ directions are
shown together with the speed distribution in the 
$y$-$z$ plane. All velocities are relative to the
mean velocity. Maxwell-Boltzmann distributions are also plotted for
comparison, with the molecular clouds having a best-fit temperature of 11~mK.

Potential developments for future molecule chips include the
integration of, for example, optical traps~\cite{Padgett:2011p1196},
as well as cooling techniques to reach ultracold temperatures and the implementation of
high-resolution spectroscopy.\cite{Carr:2009p2244} If the molecule chip were to be used as a
molecular source for high-resolution spectroscopy, the molecules would
need to be ejected at low velocity from the chip with a narrow velocity
distribution so that the packet of molecules remains together for the longest possible
time.\cite{QuinteroPerez:2013p133003} Indeed this is one of the main
goals of cold-molecule research. As a demonstration of the new
on-chip detection system, we use it to analyze the effect of a
phase-space manipulation sequence to compress the velocity
distribution without losing molecules.
With CO molecules in deep microtraps, we
slowly reduce the trap depth while guiding
the molecules at constant velocity across the chip surface.
Analysis of trajectory simulation results shows that adiabatically changing 
the trap depth from 72~mK to 13~mK in 188\microsec\ (slow compared to the trap
frequencies of hundreds of kHz) leads to a reduction of the best-fit
temperature of the molecular ensemble to roughly a third of its initial value.
Integrated imaging signals along with corresponding
trajectory simulations are shown in Fig.~\ref{fig:cooling}.
The experimental data show a clearly 
narrower spatial structure in the case of the manipulated molecules, 
indicating that, over the expansion time of 15\microsec, the manipulated molecules
have expanded less.
The in-trap spatial expansion that results from the phase-space manipulation
can be neglected for most practical purposes because the size of the
molecular clouds after release is entirely dominated by the expansion
velocity even at very short times, as can be seen in
Fig.~\ref{fig:cooling}. 
Crucially, we lose less than 5\% of the molecules during the
manipulation process. 
It is also possible to use our imaging setup for velocity
map imaging~\cite{Eppink:1997p3477}. However, with the present
ionization scheme, ion recoil energies are of 
the order of tens of mK, i.e.~comparable with the trap depths, which
makes this method imprecise for the analysis of the velocity
distributions inside the microtraps.

The chief advantage of on-chip detection is the ability to probe the
system at short evolution times. This allows for the detection of
short-lived quantum states. Moreover, it maximizes spatial 
resolution and increases the signal-to-noise ratio, as it avoids the
ballistic expansion of the molecules on their way to an external
detector. 
Here we have presented an on-chip detector that fulfills all
these requirements. One of its main advantages is its general applicability, 
through the use of the REMPI process, making it
not only quantum-state-selective, but applicable to virtually every
molecule. 
This development of the molecule chip to a complete environment
that can trap, manipulate and detect molecules is significant progress
towards a tool for the control of single and small samples of
molecules for investigating a wide range of quantum phenomena and
molecular processes.

\section*{Acknowledgments}
We gratefully acknowledge the work of the electronic laboratory of the
Fritz Haber Institute, in particular Georg 
Heyne and Viktor Platschkowski, as well as fruitful discussions with
Samuel A.~Meek. This work has been funded by the
European Community's Seventh Framework Program FP7/2007-2013 under
grant agreement 216 774 and ERC-2009-AdG under grant agreement
247142-MolChip. 


\bibliographystyle{gams-notit-nonumb}
\bibliography{coldmol}

\end{document}